\documentclass[sigconf]{acmart}

\pdfoutput=1
\usepackage[linesnumbered,commentsnumbered,ruled]{algorithm2e}
\usepackage{amsmath}
\usepackage{subfigure}
\usepackage{enumitem}
\usepackage{caption}
\usepackage{hyperref}
\usepackage{xcolor}
\usepackage{multirow}
\usepackage{balance}
\PassOptionsToPackage{hyphens}{url}
\usepackage{url}
\usepackage{diagbox}

\AtBeginDocument{%
  \providecommand\BibTeX{{%
    \normalfont B\kern-0.5em{\scshape i\kern-0.25em b}\kern-0.8em\TeX}}}

\setcopyright{acmcopyright} 

\copyrightyear{2020}
\acmYear{2020} 
\setcopyright{iw3c2w3}
\acmConference[WWW '20]{Proceedings of The Web Conference 2020}{April 20--24, 2020}{Taipei, Taiwan} 
\acmBooktitle{Proceedings of The Web Conference 2020 (WWW '20), April 20--24, 2020, Taipei, Taiwan}
\acmPrice{}
\acmDOI{10.1145/3366423.3380145}
\acmISBN{978-1-4503-7023-3/20/04}

\begin{document}

\title{paper2repo: GitHub Repository Recommendation for Academic Papers}


\author{Huajie Shao$^1$,~Dachun Sun$^1$,~Jiahao Wu$^1$, ~Zecheng Zhang$^1$,~Aston Zhang$^2$}
\author{Shuochao Yao$^1$,~Shengzhong Liu$^1$,~Tianshi Wang$^1$,~Chao Zhang$^3$,~Tarek Abdelzaher$^1$}
\affiliation{%
  \institution{$^1$Department of Computer Science, University of Illinois at Urbana Champaign \\
   $^2$Amazon Web Services Deep Learning \\
   $^3$ School of Computational Science and Engineering, Georgia Institute of Technology
   }
\institution{$^1$ \{hshao5, dsun18, jwu108, zzhan147, syao9, sl29, tianshi3, zaher\}@illinois.edu \\
$^2$ astonz@amazon.com, $^3$ chaozhang@gatech.edu
	}
}

\renewcommand{\shortauthors}{Huajie Shao, et al.}


\begin{abstract}
GitHub has become a popular social application platform, where a large number of users post their open source projects. In particular, an increasing number of researchers release repositories of source code related to their research papers in order to attract more people to follow their work. Motivated by this trend, we describe a novel item-item cross-platform recommender system, \textit{paper2repo}, that recommends relevant repositories on GitHub that match a given paper in an academic search system such as Microsoft Academic. The key challenge is to identify the similarity between an input paper and its related repositories across the two platforms, \textit{without the benefit of human labeling}. Towards that end, \textit{paper2repo} integrates text encoding and constrained graph convolutional networks (GCN) to automatically learn and map the embeddings of papers and repositories into the same space, where proximity offers the basis for recommendation. To make our method more practical in real life systems, labels used for model training are computed {\em automatically\/} from features of user actions on GitHub. In machine learning, such automatic labeling is often called {\em distant supervision\/}. To the authors' knowledge, this is the first \textit{distant-supervised cross-platform} (paper to repository) matching system. We evaluate the performance of \textit{paper2repo} on real-world data sets collected from GitHub and Microsoft Academic. Results demonstrate that it outperforms other state of the art recommendation methods. 
\end{abstract}



%



\maketitle

{
\section{Introduction}
This paper proposes a novel {\em item-item recommender system that matches papers with related code repositories\/} (on GitHub~\cite{jiang2017open}) based on a {\em joint embedding\/} of both into the same space, where shorter distances imply a higher degree of relevance. The joint embedding is novel in considering both text descriptions (of papers and repositories) as well as their relations (to other papers and repositories) expressed in appropriately defined graphs. 

The work is motivated by the growing popularity of GitHub as a platform for collaboration on open source projects and ideas. In recent years, more researchers from academia and industry have shared the source code of their research with others on GitHub. A particularly clear example of that trend in computer science is research on machine learning and data mining, where links are often provided to open source repositories in published papers. This trend is expected to grow, as it is attributed to the increasingly computation-intensive nature of modern scientific discovery. Advances in science are increasingly assisted by complex computing systems and algorithms. Reproducibility, therefore, calls for availing the research community not only of the published scientific results but also of the software responsible for producing them. In the foreseeable future, readers will frequently want to find the source code needed to repeat the experiments when searching for papers of interest on academic search systems, such as Google Scholar and Microsoft Academic\footnote{\url{https://academic.microsoft.com}}. Our new cross-platform recommender system, \textit{paper2repo}, addresses this need by automatically finding repositories relevant to a paper, even if pointers to the code were not included in the manuscript.

In the past few years, cross-platform recommender systems~\cite{zhuang2018cross,elkahky2015multi,jiang2015social} attracted a lot of attention. Cross-platform recommender systems refer to those that leverage information from multiple different platforms to recommend items for users. Existing work mainly adopts transfer learning to enrich users' preference models with auxiliary information from multiple platforms. For example, Elkahky et al.~\cite{elkahky2015multi} developed a multi-view deep learning model to jointly learn users' features from their preferred items in different domains of Microsoft service. Such prior work typically uses {\em supervised learning\/} to infer user preferences from their past behavior. In contrast, this paper designs a new item-item cross-platform recommender system, \textit{paper2repo}, that matches repositories on GitHub to papers in the academic search system, {\em without the benefit of prior access history or explicit labeling\/}. Instead, we use {\em joint embedding\/} to quickly find repositories related to papers of interest, thereby seamlessly connecting Microsoft Academic with GitHub (acquired by Microsoft in June, 2018\footnote{\url{https://news.microsoft.com/2018/06/04/microsoft-to-acquire-github-for-7-5-billion/}}).

A simple candidate solution to our matching problem might be to search for keywords from paper titles in the database of repositories on GitHub. However, it may not recommend sufficiently diverse items to users. For instance, a user who is interested in the textCNN repository\footnote{\url{https://github.com/dennybritz/cnn-text-classification-tf}} may like the Transformer repository\footnote{\url{https://github.com/tensorflow/models/tree/master/official/transformer}} as well, because both of them are related to natural language processing. Such a comprehensive set of relevant repositories is hard to infer from paper titles and, in fact, will often not be included in the text of the manuscript either.

Deep neural networks have recently been used to implement recommender systems. Among the recent advances, a \textit{semi-supervised} graph convolutional network (GCN)~\cite{ying2018graph,hamilton2017representation,xu2018powerful} has been widely used for recommendation because it can exploit both the graph structure and input features together. However, past work applied the approach to {\em single platforms only\/}. Since papers and repositories live on two different platforms, their embeddings generated by traditional GCN are {\em not\/} in the same space. Our contribution lies in extending this approach to {\em joint embedding\/} that bridges the two platforms. The embedding requires constructing appropriate graphs for papers and repositories as inputs to the GCN, as well as selecting the right text features for each. We address these challenges with the following core techniques:

\medskip
\noindent
\textbf{Context graphs:} Relations between papers can be described by a citations graph. Relations between repositories are less obvious. In this paper, we leverage the tags and user starring to connect repositories together. There exists an edge between two repositories if there are overlapped keywords between them, or if they are starred by the same users. In addition, term frequency-inverse document frequency (TF-IDF)~\cite{ramos2003using} is adopted to extract the important tags to construct the repository-repository context graph.

\medskip
\noindent
\textbf{Constrained GCN model with text encoding.} To map the embeddings of papers and repositories to the same space, we develop a joint model that incorporates a text encoding technique into the constrained GCN model based on their contents and graph structures. Since some research papers explicitly name (in the manuscript) their corresponding repositories of source code on GitHub, these papers could be used as \textit{bridge papers} to connect the two cross platforms. Accordingly, we propose a constrained GCN model to minimize the distance between the embeddings of bridge papers and their original repositories. Moreover, we leverage a text encoding technique with convolutional neural networks (CNN) to encode information from the abstracts of papers, and the descriptions and tags of repositories as input features of the constrained GCN model. As a result, the proposed constrained GCN with text encoding can jointly learn and map the embeddings of papers and repositories from the two platforms into the same space.

\medskip
We conduct experiments to evaluate the performance of the proposed \textit{paper2repo} system on real-world data sets collected from GitHub and Microsoft Academic. The evaluation results demonstrate that the proposed \textit{paper2repo} outperforms other compared recommendation methods. It can achieve about 10\% higher recommendation accuracy than other methods for the top 10 recommended candidates. To sum up, we make the following contributions to applications, methodology, and experimentation, respectively:
\begin{itemize}
\item \textit{Application:\/} We develop a novel item-item cross-platform recommender system, \textit{paper2repo}, that can automatically recommend repositories on GitHub to a query paper in the academic search system.
\item \textit{Methodology:\/} We propose a joint model that incorporates the text encoding technique into the constrained GCN model to jointly learn and map the embeddings of papers and repositories from two different platforms into the same space.
\item\textit{Experimentation:\/} Evaluation results demonstrate that the proposed \textit{paper2repo} produces better recommendations than other state of the art methods.
\end{itemize}

The rest of the paper is organized as follows. Section~\ref{sec:problem} presents the motivation, background, and formulation for our \textit{\textit{paper2repo}} problem. In Section~\ref{sec:model}, we introduce the overall architecture of the \textit{\textit{paper2repo}} recommender system. Section~\ref{sec:exp} evaluates its performance on real-world data sets. Section~\ref{sec:related} summarizes related work. Finally, we conclude the paper in Section~\ref{sec:conclusion}.

\section{Preliminaries}
\label{sec:problem}
In this section, we first introduce relevant features of GitHub, and use a real-world example to motivate our work. We then review traditional graph convolutional networks (GCNs) that constitute the analytical basis for our solution approach. Finally, we formulate the problem of GitHub repository recommendation and offer the main insight behind the adopted solution. 

\begin{figure}[!t]
\includegraphics[width=3.4 in]{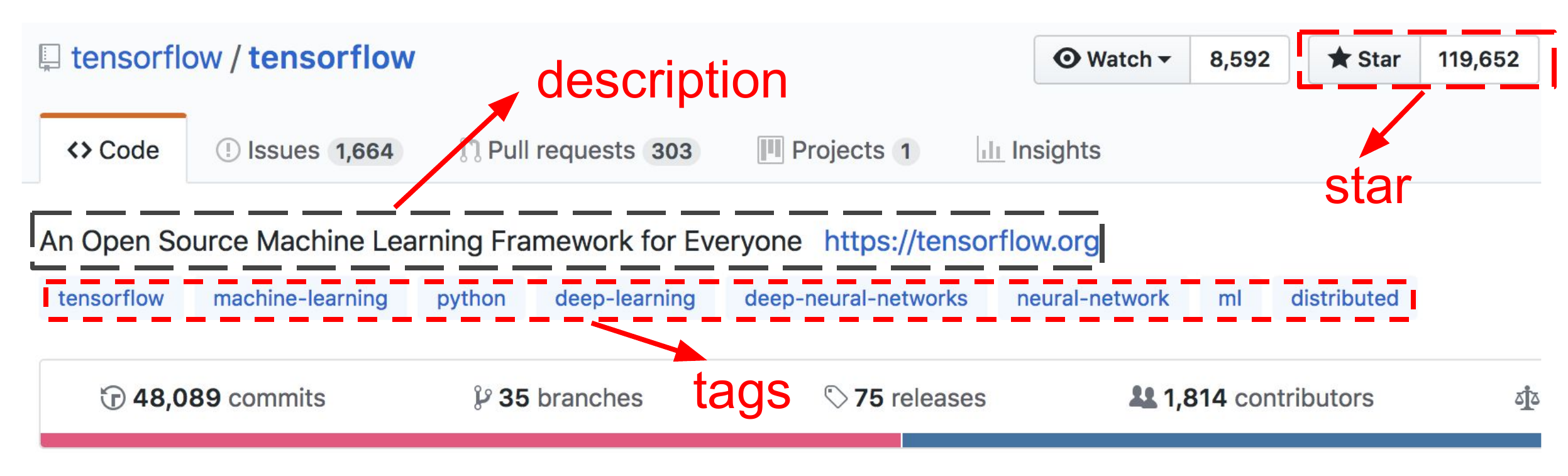}
\caption{An example of the tensorflow repository}
\label{fig:tensorflow}
\vspace{-0.1in}
\end{figure}

\subsection{Background and Motivation}
\label{sec:example}
GitHub is a social application platform, where users can collaborate on open source projects and share code with others. Users can create different repositories (\textit{i.e}., digital directories) to store files and source code for their research projects. Fig.~\ref{fig:tensorflow} illustrates an example of the Tensorflow repository for deep learning. The repository contains basic information such as descriptions, tags (also called topics), and user starring.  Descriptions refer to a couple of sentences that describe the repository. Tags are keywords used to classify a repository, including its purpose and subject area. Starring a repository represents an expression of interest from a user in the repository. Much like browser bookmarks, stars give users improved shortcut-like access to repository information.

{\color{black}
Consider a simple example to motivate features of our solution approach based on real-world data sets collected from GitHub and Microsoft Academic. Fig.~\ref{fig:repo2repo} illustrates selected repositories and papers in the two different platforms. Repository $R_1$ is the original open source code for paper $P_1$, ResNeXt~\cite{xie2017aggregated}, which presents a highly modularized network architecture for image classification developed by Facebook Research. The pair of $R_1$ and $P_1$ could be used as a bridge to connect the GitHub platform and Microsoft Academic platform. In addition, we find that users who star the repository $R_1$ also star certain other repositories related to image classification and deep neural networks. We collect 540 users who star repository $R_1$ and discover that $241$ of them (such as user $U_1$) star $R_4$ as well. 
Thus, we infer that the two repositories starred by many of the same users are likely related to each other. We can also find repositories in the same subarea or topic that have similar tags. Take repository $R_2$ and $R_3$, for example. These two repositories are related to image classification and they have the same tag ``ImageNet''. In this paper, we use both user starring and tags to construct a context graph for repositories. 

In the papers domain, related papers are connected together via citation relationships. As illustrated in Fig~\ref{fig:repo2repo}, paper $P_1$, called ResNeXt, develops a deep residual networks for image classification. It has more than 100 citing/cited papers. Most of them are also involved in the same or similar areas as $P_1$. For example, paper $P_1$ cites paper $P_2$ , $P_3$, and $P_4$ that are related to image classification using deep neural networks. Paper $P_3$ cites $P_2$ and $P_4$ because they share the same research topics on deep residual networks. Hence, related papers can be identified from the paper citation graph. We further find that the more edges (connections) exist among papers, the more related they are. These observations form the basis of constructing the context graph for publications.

Finally, we feed both the context graphs (describing relations) and content information (describing nodes), such as raw paper abstracts and repository descriptions, into the graph convolutional neural networks framework to perform joint embedding. 
}

\begin{figure}
\includegraphics[width= 3.5 in]{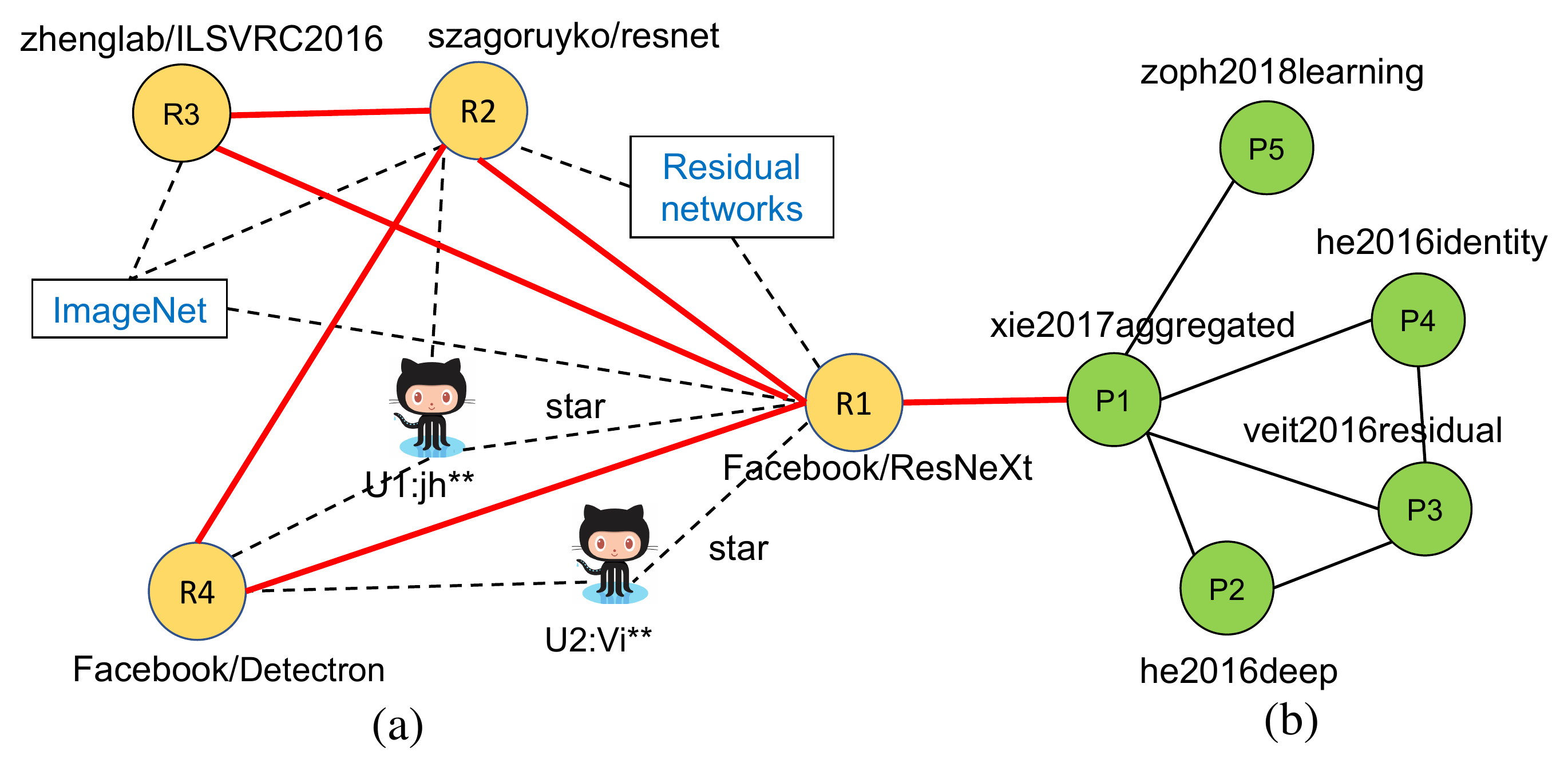}
\caption{A real-world example of repository-repository context graph on GitHub and paper citation graph on Microsoft Academic. In this figure,  each orange circle represents a repository while each green circle represent a paper. In addition, repository $R_1$ is the original open source code of paper $P_1$, which could be used to bridge the two different platforms.}
\label{fig:repo2repo}
\vspace{-0.2in}
\end{figure}

\subsection{Graph Convolutional Networks}
Graph convolutional networks (GCNs)~\cite{defferrard2016convolutional,kipf2016semi} have been widely used in classification and recommendation systems because they encode both the graph neighborhood and content information of nodes to be classified or recommended. The basic idea of GCNs is to learn the embeddings of nodes based on both their input features, $\mathbf{X}$, and graph structures. The graph structure is often denoted by an adjacency matrix, $\mathbf{A}$. A multi-layer GCN forward model could be expressed by:
\begin{equation}\label{eq:gcn}
\mathbf{H}^{(l+1)} = \sigma \Big(\tilde{\mathbf{D}}^{-\frac{1}{2}} \tilde{\mathbf{A}} \tilde{\mathbf{D}}^{-\frac{1}{2}}\mathbf{H}^{(l)} \mathbf{W}^{(l)} \Big),
\end{equation}
where $\tilde{\mathbf{A}} = \mathbf{A} + \mathbf{I}_N$, $\mathbf{A}$ is the adjacency matrix for the input graph structure and $\mathbf{I}_N$ is the identity matrix, $\tilde{\mathbf{D}}_{ii} = \sum_j \tilde{\mathbf{A}}_{ij}$, $\mathbf{W}^{(l)}$ is the weight matrix to be trained, $\sigma(.)$ is the activation function such as ReLU, $\mathbf{H}^{(l)}$ is the hidden layer, $l$ is the layer index and $\mathbf{H}^{(0)}=\mathbf{X}$ is the input features of nodes.

Because the embeddings of papers and repositories generated by traditional GCNs for the two different platforms are not in the same space, it is necessary for us to jointly train the corresponding neural networks subject to a constraint that aligns the projections of related papers and repositories onto the same (or neighboring) points in the embedding space. The above challenge leads to the following research problem formulation.

\subsection{Problem Statement and Solution Idea}
Suppose there are $N_p$ papers in an academic search system, such as Microsoft Academic, and $N_r$ repositories on GitHub. Some papers explicitly name their corresponding original repositories of source code on GitHub. We call these papers \textit{bridge papers}. They can be used to connect the two platforms. For instance, paper $P_1$ in Fig.~\ref{fig:repo2repo} is a bridge paper that connects to its original repository $R_1$ on GitHub. Each paper has an abstract, while each repository has tags and descriptions. Our main task is to learn high-quality embeddings of papers and repositories that can be used for recommendation across the two different platforms based on their content information and graph structures. 

To this end, we construct two \textit{undirected} context graphs, one for the paper platform and one for the repository platform, respectively. These graphs are introduced below:

\begin{itemize}
\item \textbf{Paper-citation graph}: We model the papers and their references as an undirected paper-citation graph, because only the connection between two papers is required in this work. We adopt an adjacency matrix with binary measurements to denote the graph structure. In this graph, each paper,  $p_i$, is a node. Each node has an abstract as its input attributes.

\item \textbf{Repository-repository context graph}: Since there is no direct citation relationship between repositories, it is more difficult to construct the repository-repository context graph. Motivated by the example in Section~\ref{sec:example}, we leverage tags and user starring to construct this graph. Specifically, as previously motivated in Fig.~\ref{fig:repo2repo}, we add a binary edge between two repositories if they are starred together by at least one user or share at least one term whose TF-IDF score is over a threshold (e.g., 0.3) in their description or tags. While binary edges ignore information available on the degree of overlap (in tags or starring users), we find that the simplification is adequate, especially that (at the end of the day) our embedding also considers full text of paper abstracts, repository descriptions, and tags.
\end{itemize}

After obtaining these two context graphs, we can combine the graph structures with the input content information of papers and repositories to generate high-quality embeddings. We can then compute similarity scores between neighboring paper and repository embeddings to recommend highly related repositories to a query paper. Namely, the top (\textit{i.e.}, nearest) repository candidates will be recommended to the query papers. Below, we elaborate this general idea in more detail.

\section{Model}
\label{sec:model}
In this section, we first introduce the overall architecture of the proposed \textit{paper2repo} recommender system, then present techniques for model training. 

\begin{figure*}[!t]
\centering
\includegraphics[width=6.35 in]{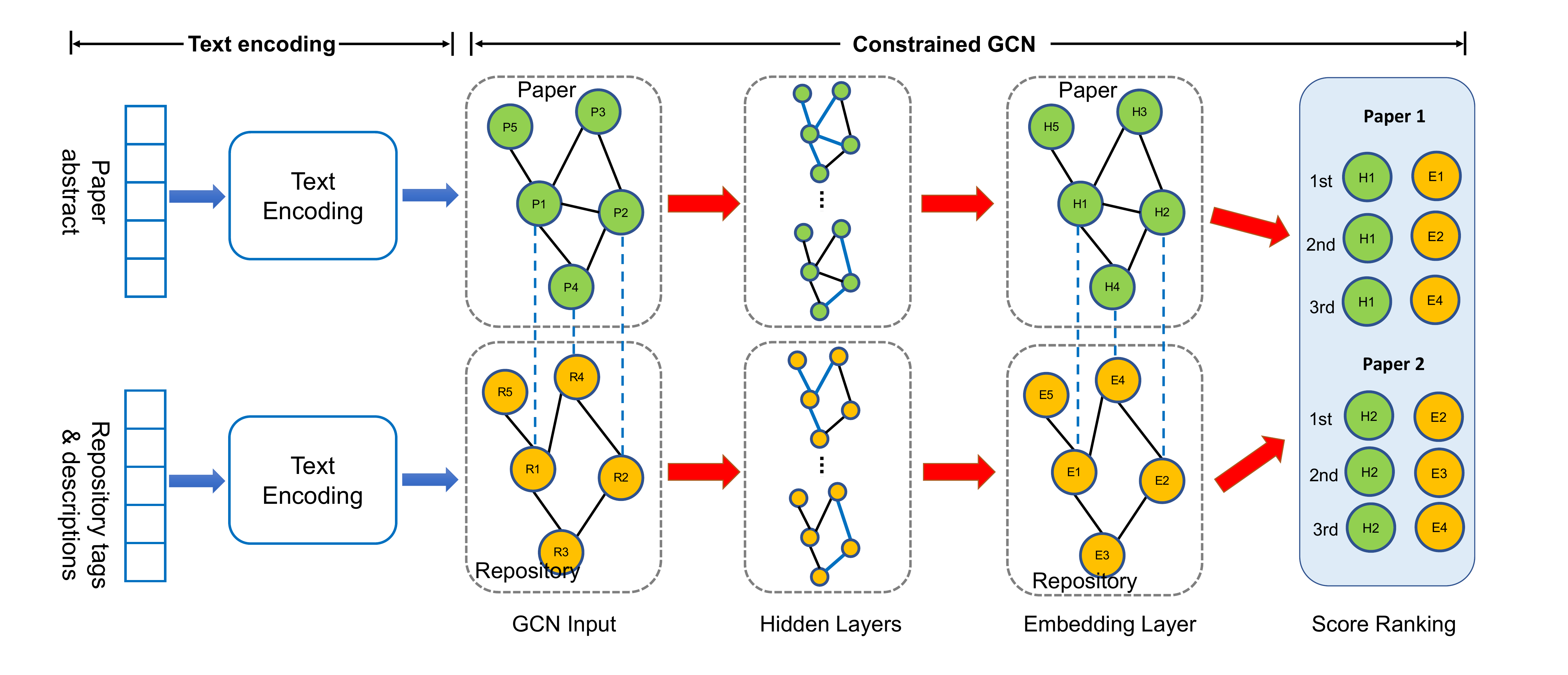}
\vspace{-0.02in}
\caption{Overall architecture of the \textit{paper2repo} system. There are two main modules: text encoding and constrained GCN model. The inputs of the \textit{paper2repo} are the content information and graph structures of papers and repositories.}
\label{fig:gcn}
\vspace{-0.02in}
\end{figure*}

\subsection{\textit{paper2repo} Architecture}
\label{sec:paper2repo}
The key contribution of this paper lies in developing a {\em joint embedding\/} that incorporates both text encoding and constrained GCN to generate embeddings that take into account similarity (between repositories and papers) in both text features and context graph structures. 

Fig.~\ref{fig:gcn} shows the overall architecture of the proposed \textit{paper2repo} system. There are two main modules: text encoding and constrained GCN. The first module tries to develop a text encoding technique to encode the content information of papers and repositories into vectors as inputs of the constrained GCN model. The second module implements the GCN model with an added constraint. The constraint specifies that repositories explicitly mentioned in paper descriptions must be mapped to the same place as the referring paper, thereby linking the mappings of repositories and papers into a single consistent space. We detail these two modules in the following.

\begin{figure}[!tp]
\centering
\includegraphics[width=3.3 in]{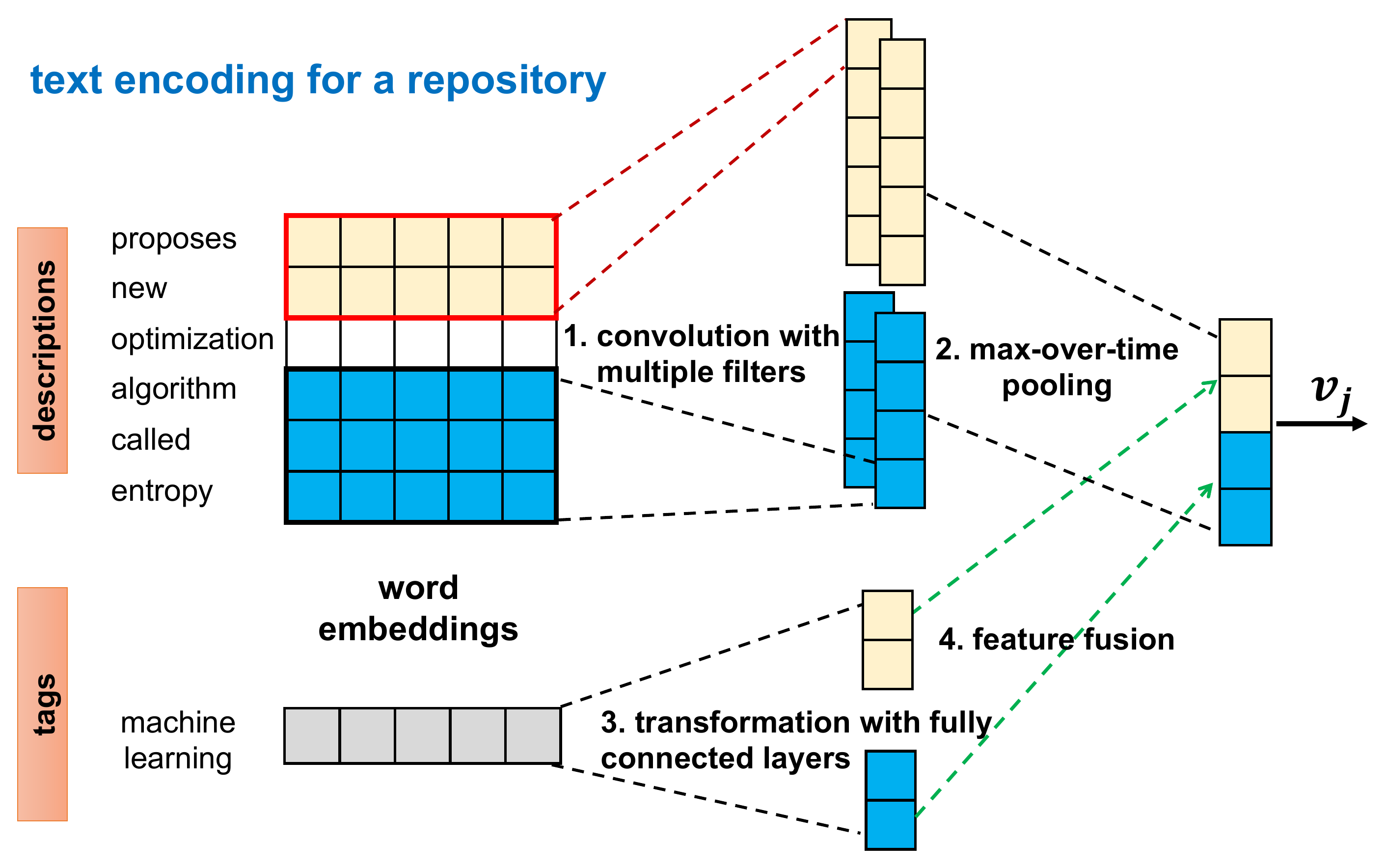}
\vspace{-0.03in}
\caption{The framework of text encoding for tags and descriptions of repositories.}
\label{fig:textCNN}
\vspace{-0.02in}
\end{figure}

\medskip
\noindent
\textbf{Text encoding.} Fig.~\ref{fig:textCNN} shows the detailed framework of using text encoding to learn the features from both the tags and descriptions of repositories on GitHub. In general, the descriptions of repositories are limited and simple, so it is difficult to learn the overall features of repositories from them. Hence, we add the tags from repositories to enrich the descriptions, because tags often extract some key information about the topics of the repositories on GitHub. Therefore, we design the text encoding module to encode the features of both tags and descriptions. As illustrated in Fig.~\ref{fig:textCNN}, there are four steps for description encoding and tag encoding of a repository:
\begin{itemize}
\item Step 1: convolution with multiple filters;
\item Step 2: max-over-time pooling;
\item Step 3: transformation with fully connected layers;
\item Step 4: feature fusion.
\end{itemize}

We first use Step 1 and Step 2 to encode the descriptions of a repository as shown in Fig.~\ref{fig:textCNN}. Let $\mathbf{x}_i \in \mathbb{R}^k$ denote the $k-$dimensional word vector of the $i$-th word in the description. The word vector comes from the pre-trained embeddings of Wiki words offered by the GloVe algorithm~\cite{pennington2014glove}. The length of each description is fixed to $n$ (padded or cropped where necessary). So a description can be defined by
\begin{equation}
\mathbf{x}_{1:n} = \mathbf{x}_1 \oplus \mathbf{x}_2 \oplus \ldots \oplus \mathbf{x}_n,
\end{equation}
where $\oplus$ is the concatenation operator that concatenates all the embeddings of words for a sentence. Then we apply filters $\mathbf{w} \in \mathbb{R}^{hk}$ for the convolution operation in order to learn new features, where $h$ is the window size of words for filters. 
Each filter computes the convolution to generate a new feature, $c_i$, from a possible window of words $\mathbf{x}_{i:i+h-1}$, which can be expressed by
\begin{equation}
c_i = f(\mathbf{w}.\mathbf{x}_{i:i+h-1} + b),
\end{equation}
where $b$ is a bias parameter and $f(.)$ is a non-linear activation function such as ReLU. As the window slides from $\mathbf{x}_{1:h} $ to $\mathbf{x}_{n-h+1:n}$, the filter yields a feature map
\begin{equation}
\mathbf{c} = [c_1,c_2, \ldots, c_{n-h+1}].
\end{equation}
After obtaining the feature map, a max-over-time pooling operation is adopted to get the maximum value, $\text{max}\{\mathbf{c}\}$, of each output feature map as illustrated in Step 2.

Next, we implement Step 3 to encode tags. Since there is no sequence for tags, we leverage fully connected layers to learn their features. For each tag, we first use the \textit{fastText} trick~\cite{joulin2016bag} to get its word representation via merging the embeddings of each word. Here, \textit{fastText} is a simple text encoding technique where word embeddings are averaged into a single vector. As illustrated in Fig.~\ref{fig:textCNN}, the embedding of ``machine learning'' would be denoted by a new word vector by adding the word vectors of ``machine'' and ``learning''. We then apply fully connected layers to produce the new features whose dimensions are aligned with the number of feature maps for description encoding. After that, feature fusion is adopted in Step 4 to add the new features of tags generated in Step 3 to the produced features in Step 2. Finally, the new features, $\mathbf{v}_j$, integrated with description encoding and tag encoding are input of the constrained GCN model. In order to improve the stability of model training, we adopt batch normalization to normalize the new features.

Similarly, for abstract encoding of a paper, we can propose the same methodology to learn the new features as the description encoding of a repository in Step 1 and Step 2. For brevity, we will not describe it in detail.

\medskip
\noindent
\textbf{Constrained GCN model.} Since papers and repositories are in two different platforms, the generated embeddings by the traditional GCN are not in the same space. Thus, we propose a constrained GCN model to constrain the embeddings to the same space. Specifically, it leverages the general GCN as the forward propagation model in Equation (\ref{eq:gcn}) and minimizes the distance between the embeddings of some bridge papers and their original repositories as a constraint. We use the cosine similarity to measure the distance. In order to compute the cosine similarity distance for the embeddings, we normalize their embeddings as $\mathbf{p}_i$ and $\mathbf{r}_j $, respectively. Let $\mathbf{p}_i^\prime$ and $\mathbf{r}_i^\prime$ denote the normalized embeddings of the $i$-th bridge paper and the corresponding original repository, respectively. Then, for each pair of bridge paper and repository, the corresponding constraint could be expressed as:
\begin{equation}\label{eq:cons}
1 - {\mathbf{p}_i^\prime}^\top {\mathbf{r}_i^\prime} \leq \epsilon,
\end{equation}
where $\epsilon$ is a small error term, such as 0.001.

In this paper, we adopt weighted approximate-rank pairwise (WARP) loss~\cite{weston2011wsabie} to train the \textit{paper2repo} model in order to recommend the target repositories to a query paper at the top of the ranked candidates. Let $\mathcal{L}$ be the WARP loss for the labeled pairs of papers and repositories during training, and $m$ be the number of pairs for bridge papers and their original repositories in the training data. Then, the constrained GCN model is defined as:
\begin{equation} \label{eq:model}
\begin{aligned}
&\text{min} \quad \mathcal{L}  \\
& \text{subject to}  &&  \sum_{i=1}^m( 1 -{\mathbf{p}_i^\prime}^\top {\mathbf{r}_i^\prime}) \leq \epsilon.
\end{aligned}
\end{equation}

\noindent
In the above~\eqref{eq:model}, the WARP loss function $\mathcal{L}$ is defined by:
\begin{equation}\label{eq:loss}
\mathcal{L} = \sum_{n_k} L \Big(rank^\Delta \big(\mathbf{p}^\top \mathbf{r}_{+} \big) \Big)  \frac{| \Delta- \mathbf{p}^\top\mathbf{r}_{+} +  \mathbf{p}^\top \mathbf{r}_{n_k} |_{+}}{rank^\Delta \big(\mathbf{p}^\top\mathbf{r}_+ \big)},
\end{equation}
where $(0<\Delta <1)$ is the margin hyper-parameter; $n_k$ denotes the number of negative examples in a batch during model training; $\mathbf{r}_{+} $ and $\mathbf{r}_{n_k}$ denote the representations of positive and negative repositories, respectively; $|t|_{+}$ is the positive part of $t$. In addition, $L(.)$ and $rank^\Delta \big(\mathbf{p}^\top \mathbf{r}_{+} \big)$ will be introduced below.

First, $L(.)$ in~\eqref{eq:loss} is a transformation function which transforms the rank into a loss, defined as
\begin{equation}
L(K)  = \sum_{j=1}^{K} \frac{1}{j},
\end{equation}
\noindent
where $K$ denotes the position of a positive example in the ranked list.
In addition, $rank^\Delta \big(\mathbf{p}^\top \mathbf{r}_{+} \big)$ is the margin-penalized rank of positive examples, defined as:
\begin{equation}
rank^\Delta \big(\mathbf{p}^\top \mathbf{r}_{+}\big) = \sum_{n_k} I\big(  \Delta- \mathbf{p}^\top \mathbf{r}_{+} +  \mathbf{p}^\top\mathbf{r}_{n_k} >0 \big),
\end{equation}
where $I(.)$ is the indicator function that outputs 1 or 0.

After defining the WARP loss,  the next step is to solve the above non-linear optimization model in~(\ref{eq:model}). One possible solution is to transform it into a dual problem. Based on the method of Lagrange multipliers, we can rewrite the above constrained model~(\ref{eq:model}) as:
\begin{equation} \label{eq:L-model}
\text{min} \quad \mathcal{L} +\lambda \sum_{i=1}^m( 1 -{\mathbf{p}_i^\prime}^\top {\mathbf{r}_i^\prime}).
\end{equation}
\noindent
It is very hard to obtain a closed-form solution that minimizes the objective function~(\ref{eq:L-model}) due to the nonlinear nature of the optimization problem. Instead, we feed this objective function as a new loss function to the neural network. Convergence of training of this neural network to a solution that meets both the original loss function, $\mathcal{L}$ (left term in objective function \ref{eq:L-model}), and the joint embedding constraint (right term in objective function \ref{eq:L-model}), requires that the two terms have comparable weights. Otherwise, learning gradients will favor optimizing the larger term and ignore the smaller. Unfortunately, the loss function $\mathcal{L}$ dynamically drops during training, making it difficult to choose the hyper-parameter, $\lambda$, to create gradients that properly trade off the loss function, $\mathcal{L} $, and the constraint error. To address this issue, we replace $\lambda$ in the second term with the varying WARP loss, $\mathcal{L}$, and normalize the constraint error instead of using the total error. Accordingly, our model can be formulated as:
\begin{equation}\label{eq:newLoss}
\text{min} \quad \big(1+ \mathcal{C}_e \big) \mathcal{L} ,
\end{equation}

\noindent
where $\mathcal{C}_e$ is the average constraint error, defined as:
\begin{equation}\label{eq:constraint}
 \mathcal{C}_e = \frac{1}{2m}\sum_{i=1}^m( 1 -{\mathbf{p}_i^\prime}^\top {\mathbf{r}_i^\prime}),
\end{equation}
Note that, $\mathcal{C}_e \in [0,1]$. This is because the cosign similarity for any two vectors, ${\mathbf{p}_i^\prime}^\top {\mathbf{r}_i^\prime}$, ranges between 1 and -1. Thus, each term in the summation above ranges between 0 and 2. The entire summation ranges between 0 and $2m$, and the normalization leads to a value of  $\mathcal{C}_e$ between 0 and 1. Using the mean constraint error, $\mathcal{C}_e$, helps keep the scale of the two terms in the objective function~(\ref{eq:newLoss}) comparable. While hypothetically, we can even remove the constant, 1, from objective function~(\ref{eq:newLoss}), we find that keeping it improves numeric stability and convergence.  

In the new formulation, we no longer need to dynamically adjust the hyper-parameter, $\lambda$, for model training. The constrained optimization problem in~(\ref{eq:model}) can be solved by minimizing the new loss function~(\ref{eq:newLoss}) by feeding it to the graph neural network.

\subsection{Model Training}
\label{sec:train}
We summarize the main notations used in our model in Table~\ref{tab:notation}. The output of our trained network is the ranked inner products of the embeddings for pairs of papers and repositories. The closer the embeddings of a paper and a repository, the higher their inner product, and the higher the ranking. Given that output, one can identify the top ranked recommended repositories for a given paper (or the top ranked papers for a given repository). Training computes the weights and biases of each convolutional neural network layer, the weights and biases of each fully connected layer (for tag encoding), and the weights and biases of filters for the convolution operation. Besides, we need to tune the number of filters used for text encoding, the filter window $h$, the number of fully connected layers for tag encoding of repositories, the number of hidden layers of neural networks, and the dimension of output embeddings of each node as well as the margin hyper-parameter for WARP loss. We set the output dimension of representations (embeddings) of each paper and repository equal in order to compute their similarity scores. We then select positive and negative samples as follows:
\begin{table}[!tp]
\centering
\caption{Main notations in our model.}
\begin{tabular}{ |l|l|}
 \hline
 Notations & Descriptions \\
 \hline
 $\mathbf{p}_i$ & $i$-th paper among all the papers \\
 $\mathbf{r}_j$ & $j$-th repositories among all the repositories \\
$\mathbf{p}_i^\prime$  & $i$-th bridge paper \\
$\mathbf{r}_i^\prime$ & bridge (original) repository of $i$-th bridge paper \\
 $\mathcal{L}$ &  WARP loss \\
 $\mathcal{C}_e$ & Mean of the constraint error\\
 $\Delta$ & Margin hyper-parameter for WARP loss \\
 $T$ & Number of top related repositories for a paper \\
\hline
\end{tabular}
\label{tab:notation}
\end{table}



\begin{itemize}
\item
{\bf Positive samples:\/} We use the bridge papers and the corresponding repositories as labeled training data. Let $p_i$ be the $i$-th paper and $r_j$ be its highly related repositories. Such pairs, $(p_i,r_j)$,  constitute positive samples. We further hypothesize that if users, who star repository A, also star repository B more often than C, then (on average) B is more related than C to repository A. Thus, besides bridge repositories, we collect the top $T$ related repositories, ranked by the frequency they are starred by users who star the original bridge repository. In order to get more training data, we also sample some one-hop neighbors of bridge papers combined with the corresponding bridge repositories to be positive examples (at most $T$). Different values of $T$ represent how liberally one interprets ``related'' (and will result in different recommender performance). We can think of this method as a form of distant supervision; an imperfect heuristic to label more samples than what can be inferred from explicit bridge references. It is important to note that we {\em do not\/} use the same heuristic for recommender {\em evaluation\/}. Rather, as we describe in a later section, we use Mechanical Turk workers to evaluate resulting recommendation accuracy, allowing us to understand the efficacy of our distant supervision framework. 
\item
{\bf Negative samples:\/}
We also introduce negative examples, referring to repositories that are not highly related to a query paper. In this work, we randomly sample $n_k$ negative examples of repositories across the entire graph to train the model. We expect the similarity scores of positive examples to be larger than those of negative examples. 
\end{itemize}

We briefly summarize the proposed constrained GCN algorithm in~\textbf{Algorithm}~\ref{alg:gcn-rank}. In this algorithm, the first $l$ layers learn the hidden vectors (embeddings) of each paper and repository using the GCN algorithm in (\ref{eq:gcn}). Then we use the normalized embeddings of paper $\mathbf{p}_i$,  and repository $\mathbf{r}_j$ to compute their similarity scores. Additionally, we try to minimize the distance between the embeddings of bridge papers and their original repositories as an added constraint.

\begin{algorithm}[!t]\small
\SetAlgoLined
\SetKwInOut{Input}{input}\SetKwInOut{Output}{output}
\Input{Features of papers and repos $\mathbf{X}_p$, $\mathbf{X}_r$, adjacency matrix $\mathbf{A}_p,\mathbf{A}_r$ for paper-citation graph and repository-repository graph}

\Output{Embeddings of paper $\mathbf{p}_i$  and repository $\mathbf{r}_j$ }
\tcc{For paper embeddings}
$\tilde{\mathbf{A}} = \mathbf{A}_p + \mathbf{I}_N$,  $\tilde{\mathbf{D}}  = \sum_j \tilde{\mathbf{A}} _{ij}$,  $\mathbf{H}^{(0)}=\mathbf{X}_p$ \;
$\mathbf{p}_i  \leftarrow \text{ReLU} \Big( \tilde{\mathbf{D}} ^{-\frac{1}{2}}\tilde{\mathbf{A}} \tilde{\mathbf{D}} ^{-\frac{1}{2}}\mathbf{H}^{(l)}\mathbf{W}^{(l)}\Big) $\;
$\mathbf{p}_i \leftarrow \mathbf{p}_i/ \|\mathbf{p}_i\|_{2} $ \;
\tcc{For repo embeddings}
$\tilde{\mathbf{A}} = \mathbf{A}_r + \mathbf{I}_N$,  $\tilde{\mathbf{D}}  = \sum_j \tilde{\mathbf{A}} _{ij}$,  $\mathbf{H}^{(0)}=\mathbf{X}_r$ \;
$\mathbf{r}_j  \leftarrow \text{ReLU} \Big( \tilde{\mathbf{D}} ^{-\frac{1}{2}}\tilde{\mathbf{A}} \tilde{\mathbf{D}} ^{-\frac{1}{2}}\mathbf{H}^{(l)}\mathbf{W}^{(l)}\Big) $\;
$\mathbf{r}_j \leftarrow \mathbf{r}_j/ \|\mathbf{r}_j \|_{2} $ \;
\tcc{Constraint for paper $\mathbf{p}_i^\prime$ and repository $\mathbf{r}_i^\prime$}
$1 - {\mathbf{p}_i^\prime}^\top \mathbf{r}_i^\prime \leq \epsilon$ \;
\tcc{Compute similarity score}
$y_{score}  \leftarrow \mathbf{p}_i^\top \mathbf{r}_j$\;
 \caption{Constrained GCN algorithm}
 \label{alg:gcn-rank}
\end{algorithm}

\section{Experiments}
\label{sec:exp}
We carry out experiments to evaluate the performance of the proposed \textit{paper2repo} on the real-world data sets collected from GitHub and Microsoft Academic. Further, we show that the proposed model performs well on larger data sets without any hyper-parameter tuning.  In addition, we conduct ablation experiments to explore how design choices impact performance. Finally, we conduct a case study to evaluate the effectiveness of the proposed method.

\subsection{Datasets}
We collected a repository data set and a research paper data set from GitHub and Microsoft Academic, respectively. Microsoft Academic provides an API for users to query the detailed entities of academic papers, such as paper title, abstract, and venue names. For a proof of concept, we query the top  20 venue names and 8 journals of computer science, such as KDD, WWW, ACL, ICML, NIPS, CVPR and TKDE, to retrieve the entities of $59,404$ raw papers from year 2010 to 2018. 
After that, we query for the titles of these papers to collect some original open source repositories through the GitHub API. We obtain about $2,427$ original repositories corresponding to the named papers. We define bridge papers as those for which we found a matching repository (which we call the bridge repository). In addition, we collect about $8,000$ popular repositories from users who star the bridge repositories. After data cleaning and preprocessing, we have $32,029$ research papers and $7,571$ repositories, including $2,107$ bridge repositories. In our experiments, we evaluate the performance of the proposed model on both small and full data sets as illustrated in Table~\ref{tab:data}.

\begin{table}[htp]
\centering
\caption{Information summary about paper and repositories data sets.}
\vspace{-0.01in}
\begin{tabular}{ |c|c|c|c|}
 \hline
Data & \# papers & \# repositories & \# bridge papers \\
\hline
Small & $11,272$ & $7,516$ & $1,386$   \\
Large & $32,029$ & $7,571$ & $2,107$\\
\hline
\end{tabular}
\label{tab:data}
\vspace{-0.03in}
\end{table}

\medskip
\noindent
\textbf{Testing and ground truth estimation.} To evaluate the accuracy of our recommender system, its results need to be compared to ground truth (that is not part of training data). For that purpose, we asked human graders on Amazon Mechanical Turk (AMT) to evaluate produced recommendations. Specifically, for each pair to be evaluated, we provided the graders with (i) the paper title, abstract, and (ii) the description, ReadMe (if available) and URL the corresponding repository on GitHub. The graders were asked to grade similarity on a three point scale: Score “2” meant that the paper and repository were highly related; score “1” meant that they were somewhat related; and score “0” meant that the pair was not related. Three graders were used per pair. After labelling, we considered those pairs that received a score of 1 or 2 from all graders to be correct matches. Pairs that received a score of 1 or 0 from all graders (and received at least one zero) were considered to be incorrect matches. Finally, pairs where graders disagreed, receiving scores 0 and 2 together (about 7\% result), were graded by an additional grader. An average was then computed. If the average score was less than 1, the pair was considered unrelated (an incorrect match). Otherwise, it was considered a correct match.

It remains to describe how we partition data into the training set and test set. We already described how positive and negative {\em training\/} pairs were selected. It is tempting to try and choose {\em test\/} papers and repositories at random. However, due to the large number of papers and repositories available on the respective platforms, this random sampling leads to a predominance of unrelated pairs, unless the test set is very large (which would be very expensive to label). Clearly, restricting testing to predominantly unrelated papers and repositories would artificially reduce the recommender's ability to find matches. In other words, the test set has to include related pairs. Hence, we first selected $580$ bridge papers and their one-hop neighbors {\em not present in the training set\/}. We then included their repositories as well as the two hop neighborhood of those repositories according to the context graph. Thus, for each paper in the test set, we included some repositories that are likely related to different degrees, and hundreds of repositories that are likely not related (i.e., repositories related to other papers). This allowed for a much richer diversity in the test set and Mechanical Turk outputs. It is important to underscore that the criteria above were used merely for paper and repository selection for inclusion in the test set. The {\em labeling\/} of the degree of match for pairs returned by the recommender was done strictly by human graders and not by machine heuristics.



\begin{figure*}[!t]
  \centering
  \subfigure[HR@K]{
  \includegraphics[width = .33\textwidth]{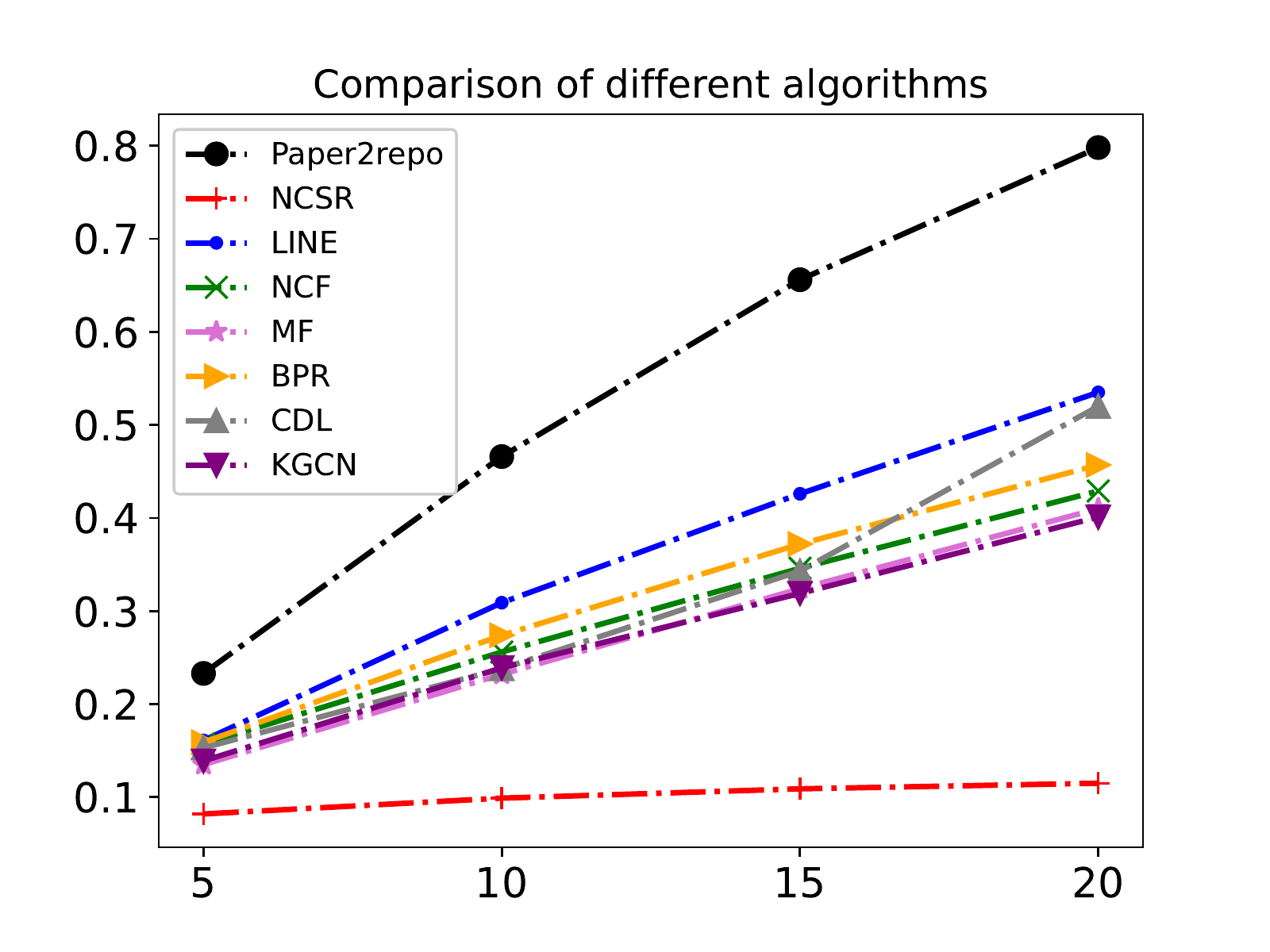}}
  \hskip -3ex
   \subfigure[MRR@K]{
  \includegraphics[width = .33\textwidth]{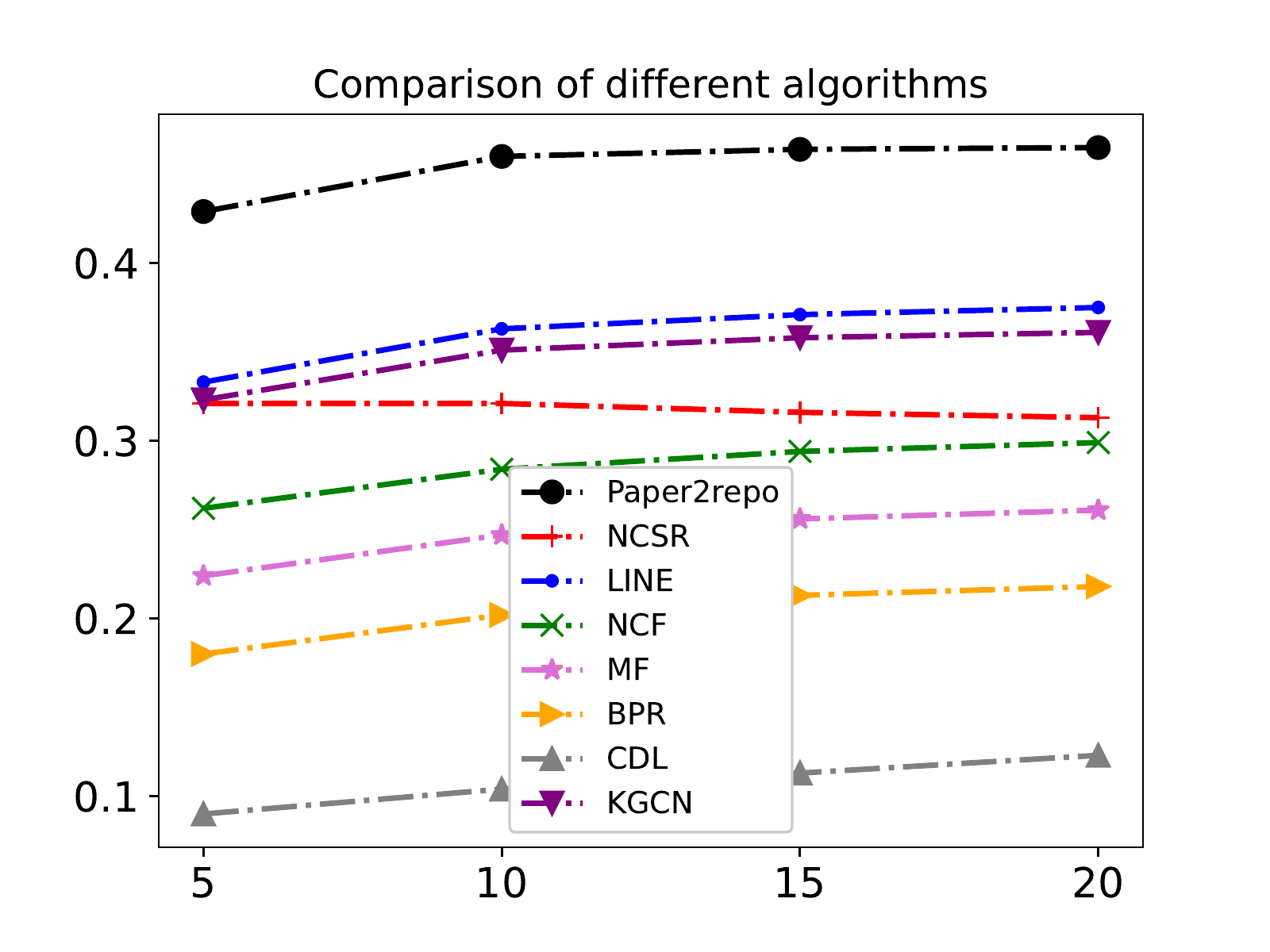}}
  \hskip -3ex
  \subfigure[MAP@K]{
  \includegraphics[width = .33\textwidth]{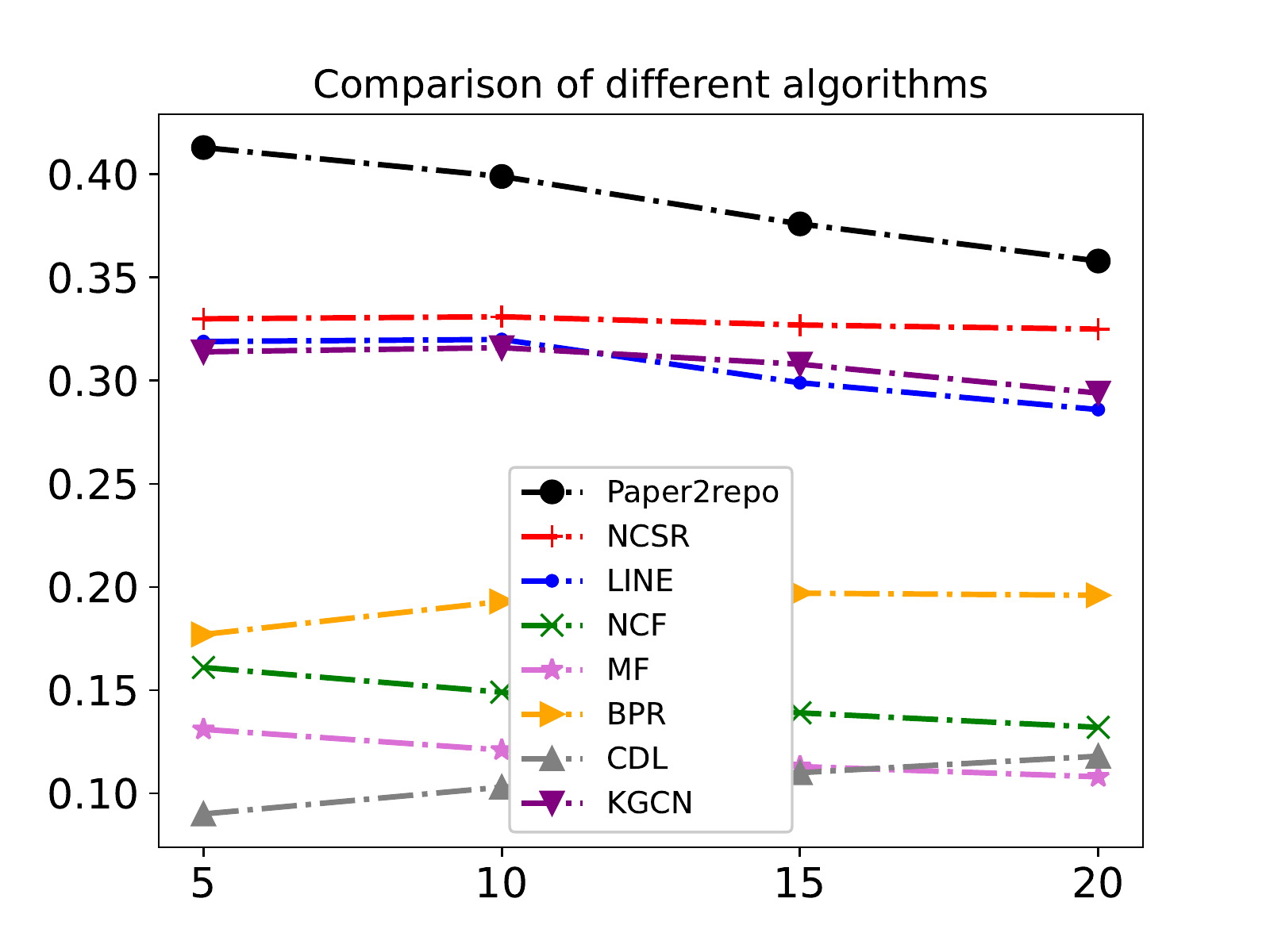}}
   \vspace{-0.12in}
  \caption{Performance comparison of different methods as the number of candidates $K$ increases.}
   \label{fig:baseline}
   \vspace{-0.1in}
\end{figure*}

\begin{figure*}[!t]
  \centering
  \subfigure[HR@K]{
  \includegraphics[width = .32\textwidth]{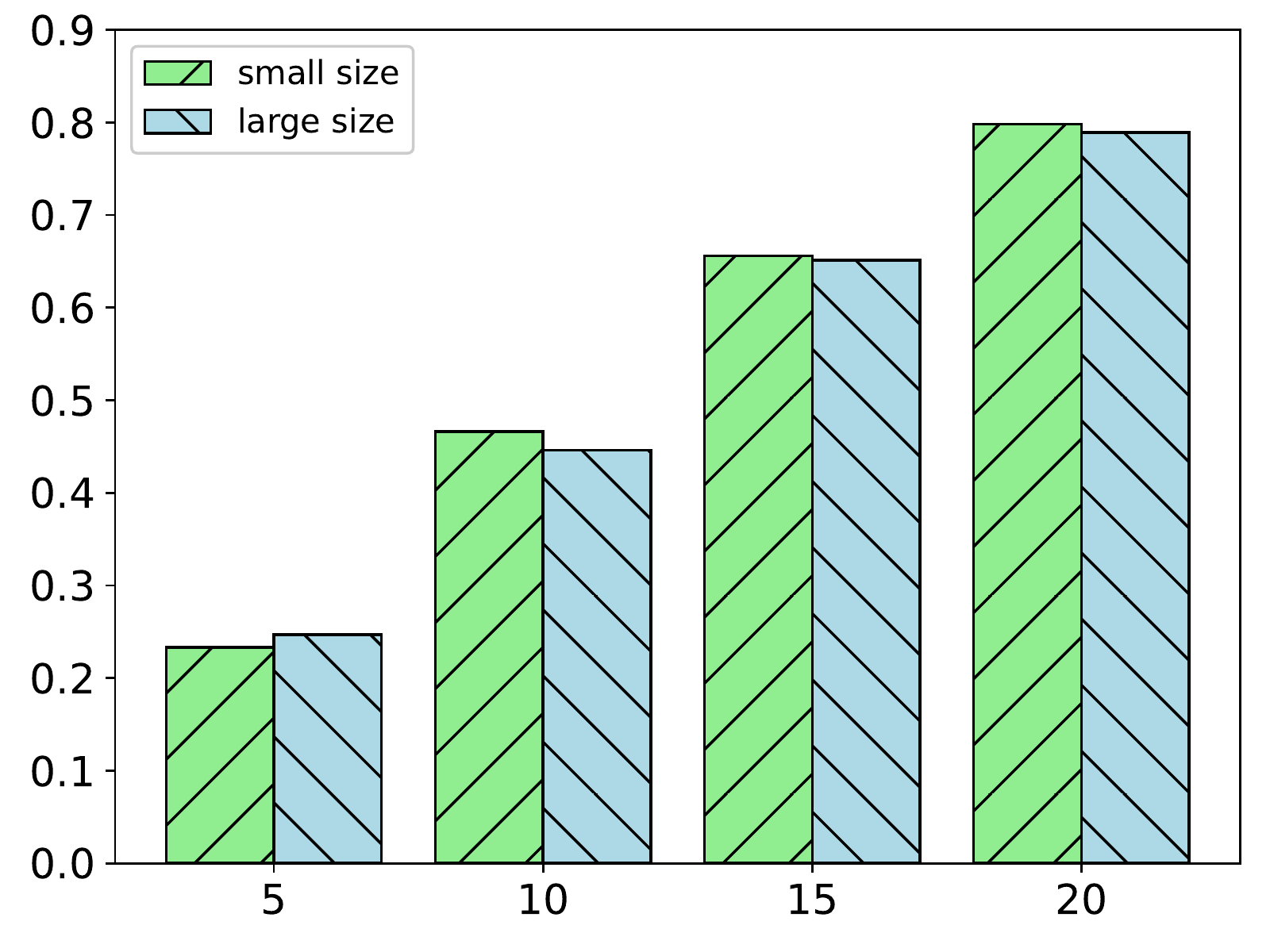}}
  \hskip -1ex
   \subfigure[MRR@K]{
  \includegraphics[width = .32\textwidth]{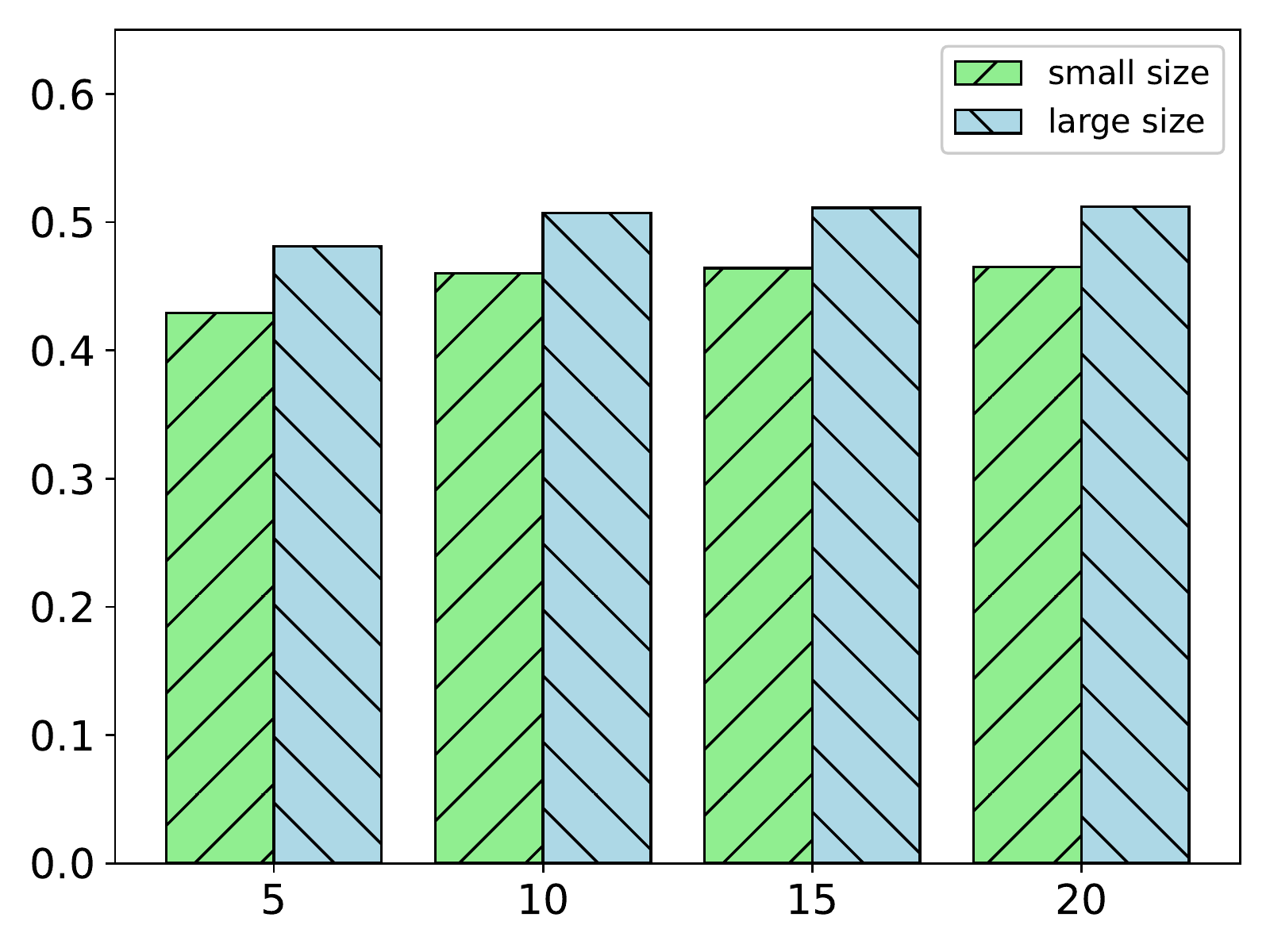}}
  \hskip -1ex
  \subfigure[MAP@K]{
  \includegraphics[width = .32\textwidth]{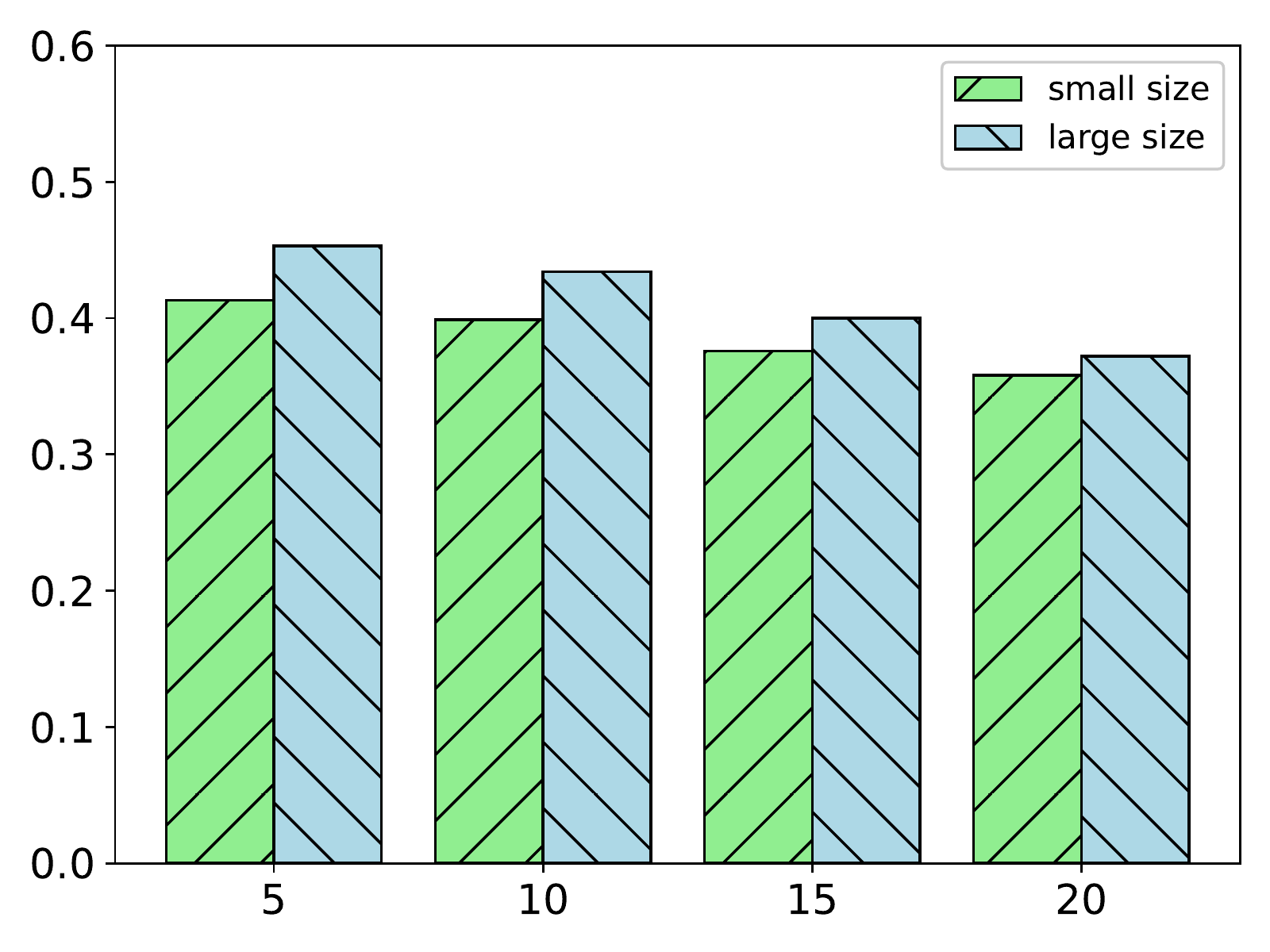}}
    \vspace{-0.07in}
  \caption{Performance comparison of different sizes of data sets using same parameters.}
   \label{fig:scalability}
   \vspace{-0.05in}
\end{figure*}

\subsection{Experimental Settings}
\textbf{Baseline methods.} We compare the proposed \textit{paper2repo} with the algorithms below:
\begin{itemize}
\item NSCR~\cite{wang2017item}: This is a cross-domain recommendation framework that combines deep fully connected layers with graph Laplacian to recommend items from information domain to social domains.
\item KGCN~\cite{wang2019knowledge}: This method leverages knowledge graph and graph convolutional neural networks to recommend interested items to users.
\item CDL~\cite{wang2015collaborative}: {\color{black}This is a hybrid recommendation algorithm, which jointly performs deep representation learning for the content information and collaborative filtering for the ratings matrix.}
\item NCF~\cite{he2017neural}: This is a neural collaborative filtering model for recommendation, which combines the matrix factorization (MF) and MLP to learn the user-item relationship.
\item LINE~\cite{tang2015line}: This is a graph embedding algorithm that uses BFS to learn the embedding of nodes in the graph with unsupervised learning. In order to better perform LINE, we construct the entire graph of papers and repositories via the bridge papers and their original repositories.
\item MF~\cite{van2013deep}: This method is widely used for traditional recommender systems due to its good performance for dense data. It is a supervised algorithm, so we use 50\% positive examples as training data and the remaining data as testing data.
\item BPR~\cite{rendle2009bpr}: This is an optimized MF model with Bayesian analysis for implicit recommendation. It is a supervised model and we use the same method as MF to do experiments.
\end{itemize}

\noindent
\textbf{Evaluation measures.} In order to evaluate the performance of the proposed \textit{paper2repo} recommender system, this paper adopts three commonly used information retrieval measures: HR@K (hit ratio), MAP@K (mean average precision), MRR@K (mean reciprocal rank). In general, HR is used to measure the accuracy about the number of correct repositories recommended by the \textit{paper2repo} system, while MAP and MRR measure the accuracy of the rank of recommended items.


\medskip
\noindent
\textbf{Model architecture and parameters.} In order to compare the performance of our proposed \textit{paper2repo} to the other recommender algorithms, we tune the hyper-parameters during model training. We use 90\% of {\em training data\/} to train the model and the remaining training data as validation data. After extensive experiments (not shown), we obtain the best parameters for our model below: we set the number of layers to 2 for the graph convolutional neural networks and the size of each layer to 256. The number of fully connected layers is 2 for tags encoding of repositories. The length of each abstract from papers, and description of repositories is fixed to 200 and 50, respectively. For paper abstract encoding, the filter window ($h$) is 2, 3, 5, 7 with 64 feature maps each. For repository descriptions encoding, its filter windows ($h$) is 2, 4 with 64 and 32 feature maps, respectively. The pre-trained embeddings of words produced by GloVe are used with size of 200. We set the learning rate to 0.0005. For training, we set the number of positive repositories, $T$, to $6$ given a paper. In addition, we randomly sample $44$ negative samples to train the model. We set the margin hyper-parameter, $\Delta$, to 0.5 in this experiment. For testing, we sample 50 examples include $T$ positive examples and the remaining $50-T$ negative example to measure the three metrics above. Our evaluation results are obtained by averaging 3 runs per point.

\subsection{Performance Comparison}
\label{sec:comp}
To understand dependence on scale, we start with a subset of $11,272$ papers and $7,516$ repositories, including $1,386$ pairs of bridge papers and original repositories, to conduct experiments (later we shall use the full data set).
We compare the performance of the proposed \textit{paper2repo} with the competing methods. Fig.~\ref{fig:baseline} illustrates the comparison results of different methods as $K$ increases with three measures: HR@K, MRR@K and MAP@K. We can observe that the proposed \textit{paper2repo} outperforms the other recommendation algorithms for all the three metrics. This is because our \textit{paper2repo} encodes both the contents and graph structures of papers and repositories to learn their embeddings. In addition, \textit{paper2repo} performs better than the cross-domain NCSR algorithm for two reasons. First of all, NCSR leverages spectral embeddings method without using node features in social domain, while our model encodes both content information and graph structures. Secondly, the items in our data sets have more attributes (keywords and tags) than that (20 attributes) in the NCSR paper, making it hard to train the NCSR model. In Fig.~\ref{fig:baseline}(c), we also find that the MAP of \textit{paper2repo} slightly drops. This reason is that more positive repositories are recommended but not at the top of ranking list, leading to lower MAP. Moreover, we can observe that CDL and NCF do not perform well in our case. The main reason is that the number of positive examples for model training is very small such that they suffer from data sparsity as the traditional MF.
Besides, KGCN does not work very well because it only adopts one hot embedding without taking node features into account in the paper domain.

\subsection{Performance on Larger Data Sets}
We conduct another experiment to evaluate the performance of the \textit{paper2repo} on larger data sets. We do not otherwise perform any data-set-specific tuning. We change the number of papers from $11,272$ to $32,029$, and the number of pairs of bridge papers and their original repositories from $1,386$ to $2,107$. Fig.~\ref{fig:scalability} illustrates the comparison results for two different sizes of data sets. From this figure, it can be observed that the proposed \textit{paper2repo} performs better on larger data sets than that on small data sets in terms of MAP and MRR, and their hit ratios are comparable. The is because our model can learn more information from different nodes in a large-scale graph.

\subsection{Ablation Studies}
\textbf{Effect of Pre-trained Embeddings}. We first explore how the settings of pre-trained embeddings impact the performance of the proposed \textit{paper2repo}. We compare three different cases: (i) fixed pre-trained embeddings; (ii) pre-trained embeddings are used as initialization (not fixed); (iii) fixed and not fixed embeddings are concatenated together. Table~\ref{tab:pre-train} illustrates the comparison results for these three cases when $K=10$. We can see that when the pre-trained embeddings are fixed, it performs better than the other two cases. The main reason is that the pre-trained embeddings are produced from the large Wikipedia corpus by GloVe. The performance of concatenated embeddings seems not very good, because it is more difficult to train the model with more complicated networks.
\begin{table}[htp]
\centering
\caption{Performance comparison for different settings of pre-trained embeddings when $K=10$, $T=6$.}
\vspace{-0.04in}
\begin{tabular}{ |c|c|c|c|}
 \hline
 Pre-trained embeddings &MAP & MRR & HR \\
 \hline
 fixed& \textbf{0.399} & \textbf{0.460} &  \textbf{0.466}  \\
not fixed & 0.330  & 0.364 & 0.367 \\
fixed \& not fixed & 0.332  & 0.374 & 0.374  \\
\hline
\end{tabular}
\label{tab:pre-train}
\vspace{-0.03in}
\end{table}

\noindent
\textbf{Effect of top $T$ positive repositories }. Next, we explore how the number of positive repositories in training, $T$, influences the performance of \textit{paper2repo}. In our experiment, the number of positive repositories, $T$, varies from 3 to 7 with step 1. From Table~\ref{tab:topT}, it can be seen that the performance of \textit{paper2repo} gradually boosts as $T$ increases from 3 to 6. When $T=6$ and $7$, their performance are close to each other. Thus, we can find that \textit{paper2repo} has a better performance when the number of positive repositories in training is large in some degree. In this paper, we use $T=6$ to do experiments because most of papers in the testing data labelled by graders have at most 6 positive examples.
\begin{table}[!tp]
\centering
\caption{Performance comparison for different $T$ when $K=10$.}
\vspace{-0.05in}
\begin{tabular}{ |c|c|c|c|c|c|}
 \hline
Top $T$ &3 & 4 & 5 & 6 &7 \\
\hline
MAP &0.269 & 0.323 & 0.350 & \textbf{0.399} & 0.388 \\
MRR &0.291 & 0.36 & 0.390 & \textbf{0.460} & 0.438 \\
HR & 0.409 &  0.414 & 0.461 & \textbf{0.466} &0.432\\
\hline
\end{tabular}
\label{tab:topT}
\vspace{-0.05in}
\end{table}

\noindent
\textbf{Effect of the Margin Hyper-parameter}. We also study the impact of the margin hyper-parameter on the performance of the \textit{paper2repo}. We change the margin parameter, $\Delta$, from 0.1 to 0.7 with step 0.1 while keeping the other hyper-parameters unchanged. Table~\ref{tab:margin} illustrates the evaluation results for different margin parameters when $K=10$. We can observe that 
the performance of \textit{paper2repo} gradually increases  as $\Delta$ rises from $0.1$ to 0.4, and then gradually drops as $\Delta$ increases from 0.4 to 0.7. This is to say, when $\Delta=0.4$, it performs best among them. The main reason is that as margin $\Delta$ is too small, it is hard to separate the positive and negative examples. At the same time, a larger $\Delta$ may result in higher loss during training.
\begin{table}[!tp]
\centering
\caption{Performance comparison for different margin parameters when $K=10$.}
\vspace{-0.05in}
\begin{tabular}{|c|c|c|c|c|c|c|c|c|c|}
 \hline
$\Delta$ &0.1&0.2 & 0.3 & 0.4&0.5 & 0.6 & 0.7 \\
\hline
MAP &0.394 &0.392 &0.394 &\textbf{0.399}& 0.349 & 0.307 &  0.274  \\
MRR &0.436 &0.456 &0.440 & \textbf{0.460}& 0.400 & 0.345 & 0.312  \\
HR    &0.446 &\textbf{0.500} &0.466 &0.466 &0.412 & 0.381 & 0.310  \\
\hline
\end{tabular}
\label{tab:margin}
\vspace{-0.05in}
\end{table}

\noindent
\textbf{Effect of Number of Bridge Papers}.
Finally, we explore how the number of bridge papers and repositories affect the recommendation performance of \textit{paper2repo}. Table~\ref{tab:bridgePaper} illustrates the comparison results under different ratios of $1,386$ bridge papers in the small data set. We can observe from it that the evaluation metrics, HR, MAP and HRR, gradually rise as the number of bridge papers increases. This is because more bridge papers and repositories can make the embeddings of the similar nodes closer to each other in the graph. 
\begin{table}[htp]
\centering
\caption{Performance comparison for different number of bridge papers when $K=10$.}
\vspace{-0.03in}
\begin{tabular}{|c|c|c|c|c|}
 \hline
Ratio &0.4&0.6 & 0.8 & 1.0 \\
\hline
MAP &0.298 &0.314 &0.341 &\textbf{0.399}  \\
MRR &0.325 &0.364 &0.391 & \textbf{0.460}  \\
HR    &0.339 &0.369 &0.379 &\textbf{ 0.466} \\
\hline
\end{tabular}
\label{tab:bridgePaper}
\vspace{-0.04in}
\end{table}


\subsection{Discussion of \textit{paper2repo}}
We further discuss the advantages and disadvantages of \textit{pape2repo}. According to our experiments, we discover that the recommended repositories are relevant to a query paper when there exists substantial user co-starring between bridge repositories and other repositories or when there are multiple overlapped tags between them. The main reason is that two repositories starred by many of the same users or that have multiple overlapped keywords are very likely to involve similar research topics. However, when only few users star both repositories, the recommendation performance is not very good. For example, when two repositories are only starred by a couple of users together, it is hard to judge whether these two repositories are similar or not.  As a result, the recommended repositories seem not to be very relevant to the query papers. The need to find a sufficient number of bridge repositories is another limitation. Extensions of this joint embedding framework to domains with no natural  analogue to bridge papers/repositories can be a good topic for future investigation.  

Cold start is one of the most important research topics in recommender systems. As we know, when users release new source code, their repositories have very little user starring in the beginning. Thus, in practice, cold start is an issue similar to lack of sufficient co-starring, discussed above. Our \textit{pape2repo} is able to partially deal with cold start because, as mentioned in Section 2.3, we construct the repository-repository graph using tags as well. Even if a repository has very few stars, we can still use its tags to construct the repository graph. Therefore, we can still recommend some repositories to the query papers, although the qualify of recommendation will be impacted.

The accuracy of our evaluation results is impacted by the accuracy of estimating ground truth. Due to budget constraints, we used three Mechanical Turk graders per item, when they coincided. Grading reliability can be improved by increasing the graders and performing grader training.

Other limitations include the general applicability of repository recommendations. While software libraries are becoming an increasingly important research enabler, some fields (including theory) are less likely to benefit from this system. The system is also less relevant to research enabled by artifacts not in the public domain. One may also argue that authors will tend to cite relevant repositories as a best practice. This is true, but our recommender system can also uncover {\em subsequently created\/} repositories that impact a given paper, such as novel implementations of relevant libraries, or algorithm implementations on new hardware platforms and accelerators. This is the same reason why one might not exclusively rely on citations in a paper to find other relevant papers, especially if the paper in question is a few years old.

\section{Related Work}
\label{sec:related}
This section reviews related work on recommender systems that use deep learning and graph neural networks, especially for cross-domain recommendations.

%

\subsection{Cross-domain Recommendation}
In order to deal with data sparsity and cold start issues in recommender systems, recent research proposed cross-domain recommendations~\cite{elkahky2015multi,wang2017item,jiang2015social,jiang2016little} to enrich user preferences with auxiliary information from other platforms. Cross-domain recommendations get rich information on items that users prefer buying based on their history on multiple platforms. For example, some users like to purchase furniture on Walmart and buy clothes on Amazon. So we can recommend furniture to the users on Amazon next time, which can improve the diversity of recommendations. Motivated by this observation, a multi-view deep learning model was proposed by Elkahky \textit{et al.}~\cite{elkahky2015multi} to jointly learn the users' features from their past preferred items on different domains. Following this work, Lian \textit{et al.}~ \cite{lian2017cccfnet} further developed a CCCFNet model that integrates collaborative filtering and content-based filtering in a unified framework to address the data sparsity problem. However, these algorithms mainly adopt the idea of transfer learning to boost users' preferences via jointly learning users' items on multiple platforms or domains. In addition, Wang \textit{et al.}~\cite{wang2017item} proposed a user-item cross-domain framework, NSCR, to recommend items from information domains to users on social media. However, a limitation is that NSCR is a supervised learning algorithm. In contrast, we propose a item-item cross-domain recommendation framework with a distant-supervised learning model (without human labelling for training).

\subsection{Recommendation with Graph Neural Networks}
In recent years, graph neural networks have been widely proposed in recommender systems~\cite{ying2018graph,fan2019graph,chen2019icient,park2019estimating,wang2019kgat,you2019hierarchical} because they can encode both node information and graph structure. For instance, Fan \textit{et al.}~\cite{fan2019graph} developed a new framework, called GraphRec, to jointly capture interactions and opinions in the user-item graph. In addition, Shaohua \textit{et al.}~\cite{fan2019metapath} proposed a meta-path guided heterogeneous Graph Neural Network for intent recommendation on an e-commerce platform. However, these efforts apply graph neural networks to recommender sytems on a single platform.

Different from prior work, we propose a novel item-item cross-domain recommendation framework that automatically recommends related repositories on GitHub to a query paper in an academic search system, helping users find their repositories of interest quickly by a \textit{joint embeddings across platforms}.


\section{Conclusions and Future Work}
\label{sec:conclusion}
This paper developed an item-item cross-platform recommender system, \textit{paper2repo} that can automatically recommend repositories on GitHub matching a specified paper in the academic search system. We proposed a joint model that incorporates a text encoding technique into a constrained GCN formulation to generate joint embeddings of papers and repositories from the two different platforms. Specifically, the text encoding technique was leveraged to learn sequence information from paper abstracts and descriptions/tags of repositories. In order to map the representations of papers and repositories onto the same space, we adopted a constrained GCN model that forces the embeddings of bridge papers and their corresponding repositories to be equal as a constraint. Finally, we conducted experiments to evaluate the performance of \textit{paper2repo} on real-world data sets. Our evaluation results demonstrated that the proposed \textit{paper2repo} systems can achieve a higher recommendation accuracy than prior methods. In the future, we can extend our method to consider other entities (such as venues and authors), in addition to papers, to construct the knowledge graph. We can improve text embedding (e.g., by use of phrase embedding instead of word embedding). We can also investigate generalizations of our joint embedding, especially to domains with no analogue to bridge papers to serve as mapping constraints.

\section{Acknowledgments}
Research reported in this paper was sponsored in part by DARPA award W911NF-17-C-0099, DTRA award HDTRA1-18-1-0026, and the Army Research Laboratory under Cooperative Agreements W911NF-09-2-0053 and W911NF-17-2-0196. 
}

%
%


%


\balance
\bibliographystyle{ACM-Reference-Format}
\bibliography{bibliography.bib}


%
%

%

%
%

\end{document}